\documentclass{article}

\usepackage{authblk,amsmath,amssymb,graphicx,geometry,cite,color,mathptmx,courier,textcomp,verbatim}
\usepackage[table]{xcolor}

\begin{document}

\title{Cascaded half-harmonic generation of femtosecond \\frequency combs in mid-IR}

\author[1,*]{Alireza Marandi}
\author{Kirk A. Ingold}
\author{Marc Jankowski}
\author{\\Robert L. Byer}

\affil{E. L. Ginzton Laboratory, Stanford University, Stanford, CA 94305}

\affil[*]{Corresponding author: marandi@stanford.edu}
\maketitle

{\bf For the growing demand of frequency combs in mid-infrared (mid-IR), known as the ``molecular fingerprint'' region of the spectrum \cite{review}, down conversion of near-IR frequency combs through half-harmonic generation offers numerous benefits including high conversion efficiency and intrinsic phase and frequency locking to the near-IR pump \cite{fcopo}. Hence cascaded half-harmonic generation promises a simple path towards extending the wavelength coverage of stable frequency combs. Here, we report a two-octave down-conversion of a frequency comb around 1 $\mu$m through cascaded half-harmonic generation with $\sim$64\% efficiency in the first stage, and $\sim$18\% in the second stage. We obtain broadband intrinsically-frequency-locked frequency combs with $\sim$50-fs pulses at $\sim$2 $\mu$m and $\sim$110-fs pulses at $\sim$4 $\mu$m. These results indicate the effectiveness of half-harmonic generation as a universal tool for efficient phase- and frequency-locked down-conversion, which can be beneficial for numerous applications requiring long-wavelength coherent sources. \bf}

Femtosecond frequency combs in mid-IR region of the optical spectrum \cite{review} are beneficial for numerous applications ranging from molecular spectroscopy in medicine \cite{medicine} to higher harmonic generation \cite{hhg}, and dielectric laser accelerators \cite{dla}.  The most effective path for generating femtosecond mid-IR frequency combs so far has been based on nonlinear methods to down-convert or extend the well-developed near-IR frequency combs of the mode-locked lasers. Parametric down conversion in optical parametric oscillators (OPOs) \cite{opocombs}, difference frequency generation \cite{dfg} and supercontinuum generation (SCG) \cite{scg} are among the widely used methods in this category. 

Recently, progress has been made towards more direct generation techniques without using near-IR frequency combs. Quantum cascade lasers (QCLs) are shown to be effective in generation of mid-IR frequency combs with outputs similar to frequency modulated lasers \cite{qclcomb}. High-Q monolithic resonators pumped by c.w. mid-IR QCLs  have also produced frequency combs \cite{kerrqcl1, kerrqcl2}. These techniques promise a path for on-chip implementation, however, they currently have limited conversion efficiencies and spectral coverage in the mid-IR, and their frequency stabilization and temporal behavior are yet to be fully explored. Substantial research is also focused on extending the operation of femtosecond mode-locked lasers to longer wavelengths in the mid-IR, with demonstrations in free-space cavities at around 2.4 $\mu$m \cite{crzns}, and in fiber cavities at around 2.8 $\mu$m \cite{erfl}.

Half-harmonic generation --the reverse of second-harmonic generation (SHG)-- is the down-conversion of an optical input by one octave. It has been achieved in OPOs operating at degeneracy in c.w. \cite{nabors} and femtosecond \cite{sam} regimes, and has been effective in enhancing the sensitivity of spectroscopy \cite{spectroscopy}. Operation at degeneracy is particularly promising because of the broad instantaneous bandwidth of the output \cite{opgaas}, the potential for high conversion efficiency, and more importantly intrinsic phase and frequency locking of the half-harmonic output to the pump due to the phase-sensitive nature of the parametric gain at degeneracy \cite{fcopo}. In contrast to non-degenerate OPOs, either in the singly-resonant \cite{junyeopo} or the doubly-resonant configuration \cite{fcdro}, the intrinsic frequency locking at degeneracy eliminates the need to employ extra frequency stabilization systems, which can be challenging to achieve in mid-IR. Cascading multiple stages of half-harmonic generation can lead to frequency- and phase-stable frequency combs centered at multiple octaves below the pump with only one frequency stabilization system in a convenient wavelength range, which is most likely around the pump.

\begin{figure}[htbp]
\centering
\includegraphics[width=.75\textwidth]{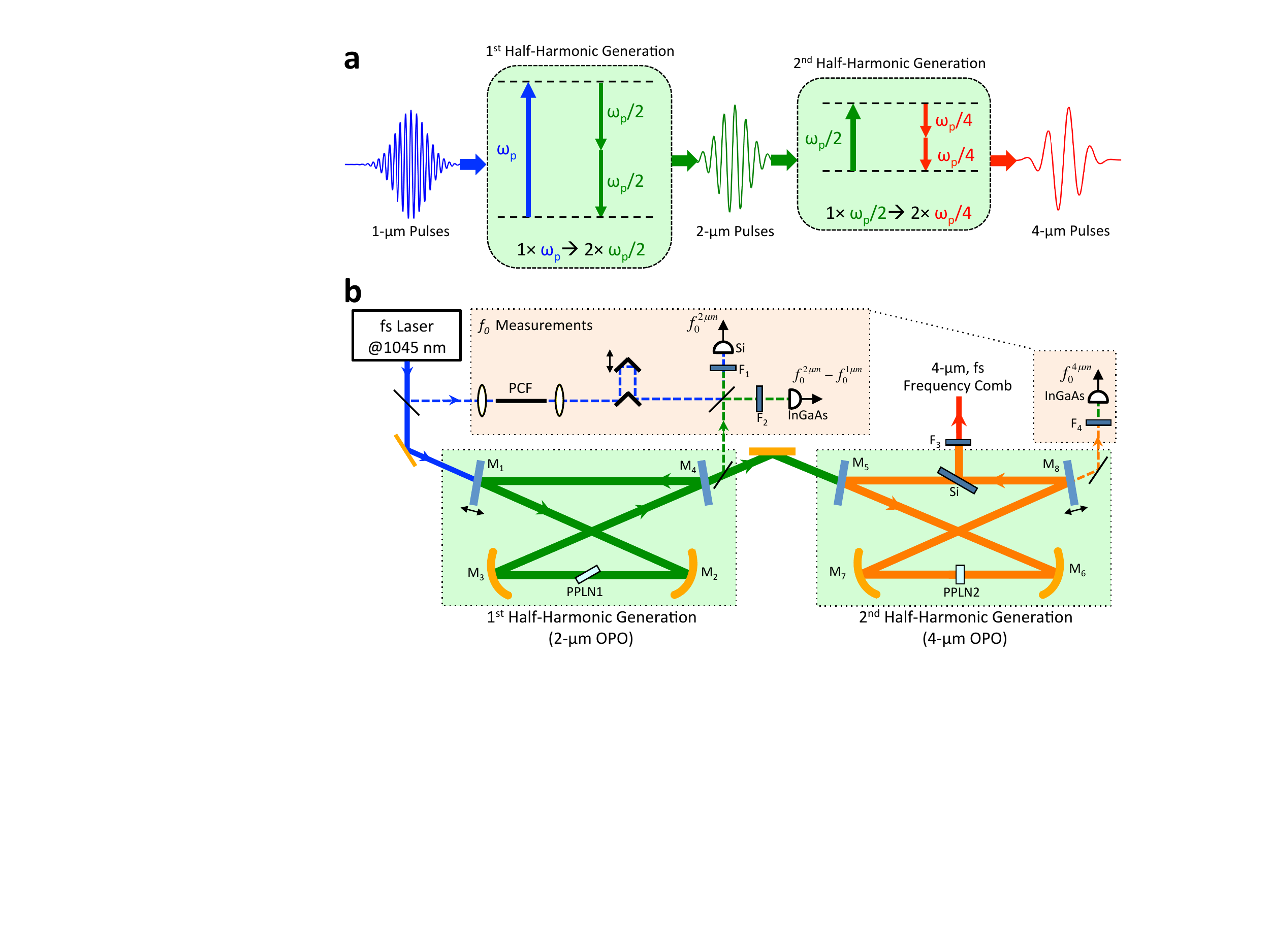}
\caption{ Cascaded half-harmonic generation of frequency combs. {\bf a,} Femtosecond pulses of the pump laser at $\omega_p$ are converted to $\omega_p/2$ and $\omega_p/4$ through two stages of intrinsically-frequency-locked half-harmonic generation. {\bf b,} Schematic of the experiment comprising two half-harmonic OPOs along with the frequency measurement setups. }
\label{fig:sch} 
\end{figure}

We cascade two half-harmonic OPOs and demonstrate efficient generation of femtosecond mid-IR frequency combs that are intrinsically frequency locked to the pump. Figure \ref{fig:sch}a illustrates the concept of cascading half-harmonic generation of femtosecond pulses. The schematic of the experimental setup is depicted in Fig. \ref{fig:sch}b. We start from a commercial mode-locked Yb:fiber laser at 1.045 $\mu$m, for which the frequency comb techniques are well-established \cite{ybcomb}, and use commercially-available periodically poled lithium niobite (PPLN) as the nonlinear crystals for the OPOs. We optimize the first half-harmonic OPO and demonstrate a record conversion efficiency of $\sim$64\% with $\sim$50-fs pulses centered at 2.09 $\mu$m. In the second half-harmonic OPO we achieve $\sim$18\% of conversion efficiency, and $\sim$110-fs pulses centered at 4.18 $\mu$m and confirm the frequency-locked operation.

\begin{figure}[htbp]
\centering
\includegraphics[width=.75\textwidth]{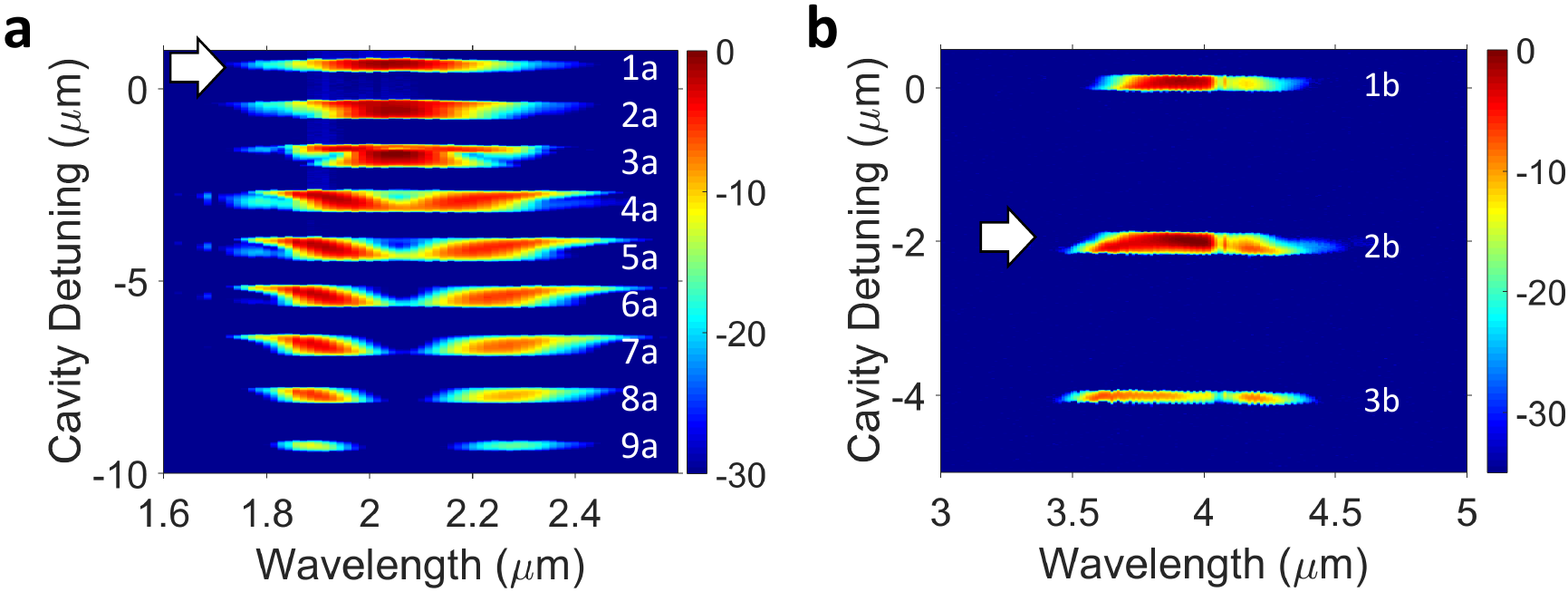}
\caption{ Measured output spectra as a function of cavity lengths for the {\bf a}, 2-$\mu$m OPO and {\bf b}, 4-$\mu$m OPO. The white arrows show the cavity lengths for stable half-harmonic generation.}
\label{fig:3d_spectrum}
\end{figure} 

The cascaded half-harmonic OPOs are pumped by a 1-$\mu$m mode-locked fiber laser (Menlo Systems Orange-A) with $\sim$70-fs pulses at 250 MHz and an average power of 950 mW. Each stage is a synchronously pumped OPO operating at degeneracy with similar bowtie resonators and roundtrip times of 4 ns matching the pump repetition period. The first OPO has an input dielectric-coated mirror (M$_1$) that is highly transmissive for the pump and highly reflective for the first half-harmonic signal around 2 $\mu m$. Two concave gold mirrors (M$_2$ and M$_3$) have a radius of curvature (ROC) of 38 mm, and the output coupler is a dielectric mirror (M$_4$) with 65\% transmission for the signal. This value is chosen based on experimental optimization of the output coupling using a 200-$\mu$m thick Ge substrate \cite{cleo2014}. The broadband parametric gain around 2 $\mu$m is provided with a 1-mm long MgO:PPLN crystal (PPLN1) with a poling period of 31.8 $\mu m$ for type-0 phase matching at room temperature, which is cut and placed at the Brewster angle. 

The second OPO is pumped by the signal output of the first OPO. It comprises two dielectric-coated flat mirrors (M$_5$ and M$_8$), which are highly transmissive around 2 $\mu$m and highly reflective around the 4 $\mu$m. Two concave gold mirrors (M$_6$ and M$_7$) have an ROC of 25 mm. The nonlinear gain is provided by a plane parallel MgO:PPLN (PPLN2) crystal with a poling period of 28.3 $\mu m$ for type-0 phase matching at room temperature, and has anti-reflection coatings for the signal and pump wavelength ranges. A 5-mm thick uncoated silicon substrate is placed in the cavity with an angle slightly away from the Brewster angle to provide output coupling of the mid-IR signal pulses and also partial intracavity dispersion compensation. The 4-$\mu$m OPO is contained in a box purged with nitrogen to reduce the effects of atmospheric absorption on the OPO operation. Its position with respect to the 2-$\mu$m OPO is chosen to result in close-to-optimal spatial overlap of the resonator modes without requiring extra mode matching optics. The threshold of the 2-$\mu$m OPO is $\sim$260 mW and the threshold of the 4-$\mu$m OPO is $\sim$250 mW.

To achieve stable half-harmonic generation from these OPOs, the roundtrip phases of their resonators have to be consistent with the phase-sensitive parametric gains. Therefore, their operation and output spectra are cavity-length dependent. The output spectrum of the 2-$\mu$m OPO as a function of its cavity length is depicted in Fig. \ref{fig:3d_spectrum}a, showing discrete oscillation peaks separated by about 1-$\mu$m length steps as a characteristic behavior of doubly-resonant OPOs. Two of these oscillation peaks, marked as peaks ``1a'' and ``2a'', correspond to half-harmonic generation, where the output spectrum has a single peak around degeneracy, i.e. at 2.09 $\mu$m. Even though the spectrum of peak ``3a'' looks degenerate, its strong cavity length dependence suggests oscillatory behaviors \cite{wolf}, which is verified experimentally. The resonator of the 2-$\mu$m OPO is stabilized to the top of the peak ``1a'' using a ``dither-and-lock'' servo controller \cite{fcopo}. The 4-$\mu$m OPO is pumped by the stable output of the 2-$\mu$m OPO at degeneracy, and its output spectrum as a function of its cavity length is depicted in Fig. \ref{fig:3d_spectrum}b. The cavity length of the 4-$\mu$m OPO is stabilized to the center of its strongest degenerate peak, marked as ``2b'' in the figure. Both OPOs could run continuously at degeneracy for several hours.

\begin{figure}[htbp]
\centering
\includegraphics[width=.75\textwidth]{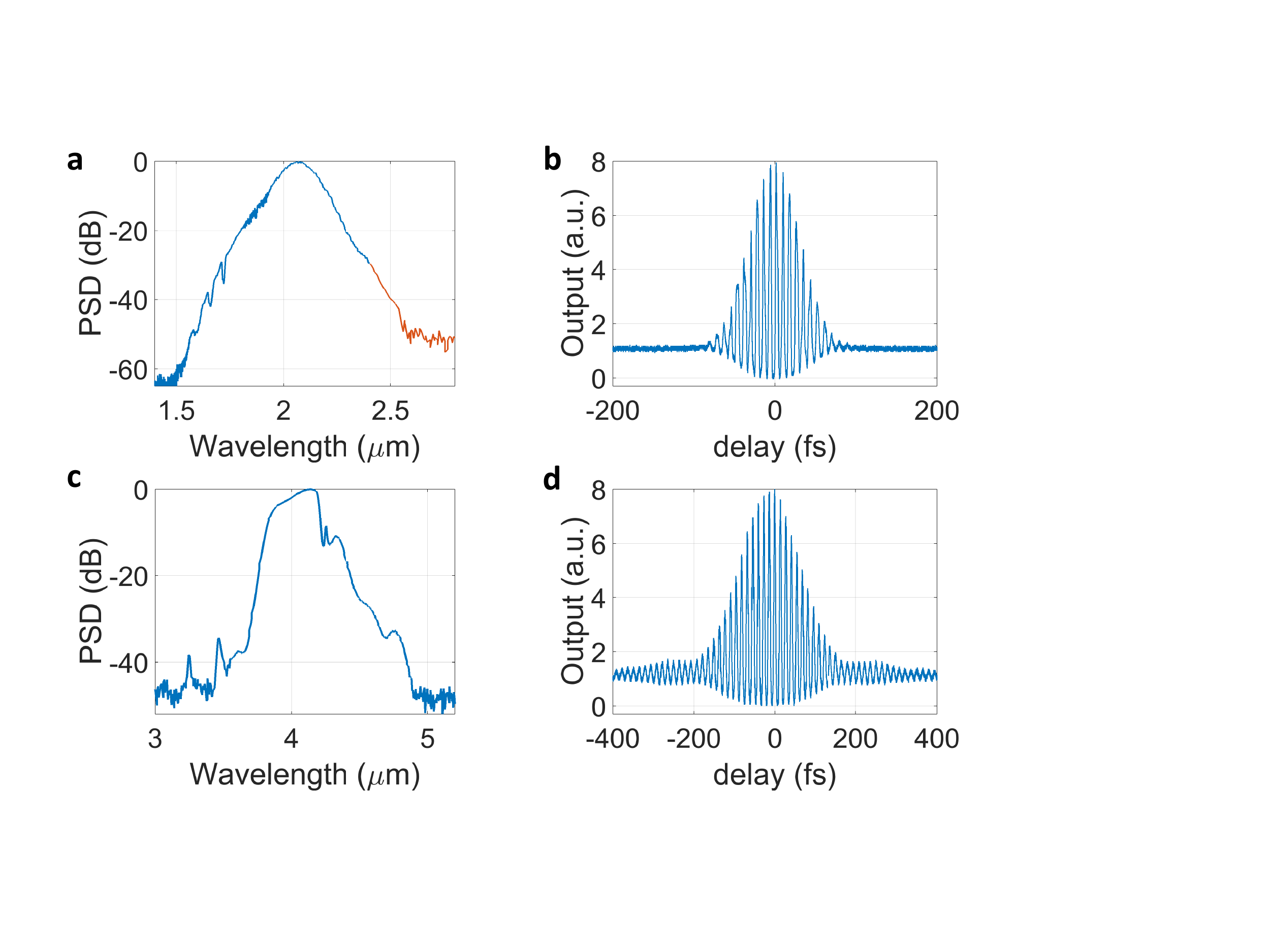}
\caption{ {\bf a}, Optical spectrum and {\bf b}, autocorrelation trace of the of the 2-$\mu$m frequency comb. {\bf c}, Optical spectrum and {\bf d}, autocorrelation trace of the of the 4-$\mu$m frequency comb. Up to 2.4 $\mu$m, the spectra are measured by a Yokogawa optical spectrum analyzer, and a 1/4-m monochromator and an InSb detector are used for above 2.4 $\mu$m. PSD is power spectral density.}
\label{fig:spectrum}
\end{figure} 

The 2-$\mu$m OPO produces $\sim$605 mW of average power with pulses centered at 2.09 $\mu m$, and its output spectrum is shown in Fig. \ref{fig:spectrum}a, which has a 3-dB bandwidth of $\sim$170 nm. The interferometric autocorrelation of the 2-$\mu$m pulses are measured by a two-photon photodetector, which is shown in Fig. \ref{fig:spectrum}b; it suggests a Gaussian pulse-width of $\sim$50 fs, which is very close to the transform limit.

The output spectrum of the 4-$\mu$m OPO is shown in Fig. \ref{fig:spectrum}c. It is centered at 4.18 $\mu m$ with the 3-dB bandwidth of $\sim$265 nm, and the 30-dB width of $\sim$1 $\mu m$. The sharp drop of the spectrum around 4.2 $\mu m$, which is limiting the 3-dB bandwidth on the red side of the spectrum, corresponds to the CO$_2$ absorption from the residual atmosphere in the purged OPO cavity and in the beam path from the OPO to the spectrometer. The autocorrelation of the pulses is measured by a two-photon detector and is shown in Fig. \ref{fig:spectrum}d. The slight chirp in the 4-$\mu$m pulses is either due to the residual dispersion in the OPO cavity or the 1-mm thick Ge substrate of the long-pass filter. The pulse-width is estimated to be $\sim$110 fs, assuming Gaussian pulses, while the transform-limited pulse length for the measured spectral width is $\sim$ 100 fs.  We expect that much shorter pulses can be achieved by minimizing the intracavity group delay dispersion and further elimination of CO$_2$ in the cavity and the measurement setup. Also, an appropriate dielectric coating on M$_8$ can be used to provide optimum out-coupling and avoid using the silicon substrate as the output coupler.

We examine the offset frequencies of the frequency combs and verify the intrinsic frequency locking of the cascaded half-harmonic generation scheme. The frequency comb lines of the pump laser, the 2-$\mu$m comb, and the 4-$\mu$m comb can be represented by:
\begin{equation}
f_\ell^{1 \mu m}= \ell f_R+ f_0^{1 \mu m},
\end{equation}
\begin{equation}
f_m^{2 \mu m}= mf_R+f_0^{2 \mu m},
\end{equation}
\begin{equation}
f_n^{4 \mu m}= nf_R+f_0^{4 \mu m},
\end{equation}
where $m$, $n$, and $\ell$ are integers, $f_R$ is the repetition frequency, and $f_0^{1 \mu m}$, $f_0^{2 \mu m}$, and $f_0^{4 \mu m}$ are the carrier envelope offset (CEO) frequencies of the frequency combs at 1 $\mu$m, 2 $\mu$m, and 4 $\mu$m, respectively. At degeneracy, the CEO frequency of the 2-$\mu$m comb is expected to be locked to the 1-$\mu$m pump by \cite{fcopo}:
\begin{equation}
f_0^{2\mu m}=\frac{f_0^{1\mu m}}{2} \quad \text{or} \quad  f_0^{2\mu m}=\frac{f_0^{1\mu m}}{2} + \frac{f_R}{2},
\label{eq:2um}
\end{equation}
and the 4-$\mu$m comb would be locked to the 2-$\mu$m comb as:
\begin{equation}
f_0^{4\mu m}=\frac{f_0^{2\mu m}}{2} \quad \text{or} \quad  f_0^{4\mu m}=\frac{f_0^{2\mu m}}{2} + \frac{f_R}{2}.
\end{equation}
The $\frac{f_R}{2}$ steps in the CEO frequencies are determined by the OPO cavity lengths, i.e. if the OPO cavity is changed from one oscillation peak of Fig. \ref{fig:3d_spectrum} to an adjacent peak, the CEO frequency will change by $\frac{f_R}{2}$. These relations do not hold for non-degenerate OPO operation \cite{fcdro}, for instance in peaks ``7a" to ``9a" of the 2-$\mu$m OPO (Fig. \ref{fig:3d_spectrum}a), or for cavity lengths supporting transitions between degenerate and non-degenerate operation, for instance peaks ``3a'' to ``6a'' of the 2-$\mu$m OPO and peak ``3b'' of the 4$\mu$m OPO (Fig. \ref{fig:3d_spectrum}a,b).

\begin{figure}[htbp]
\centering
\includegraphics[width=.75\textwidth]{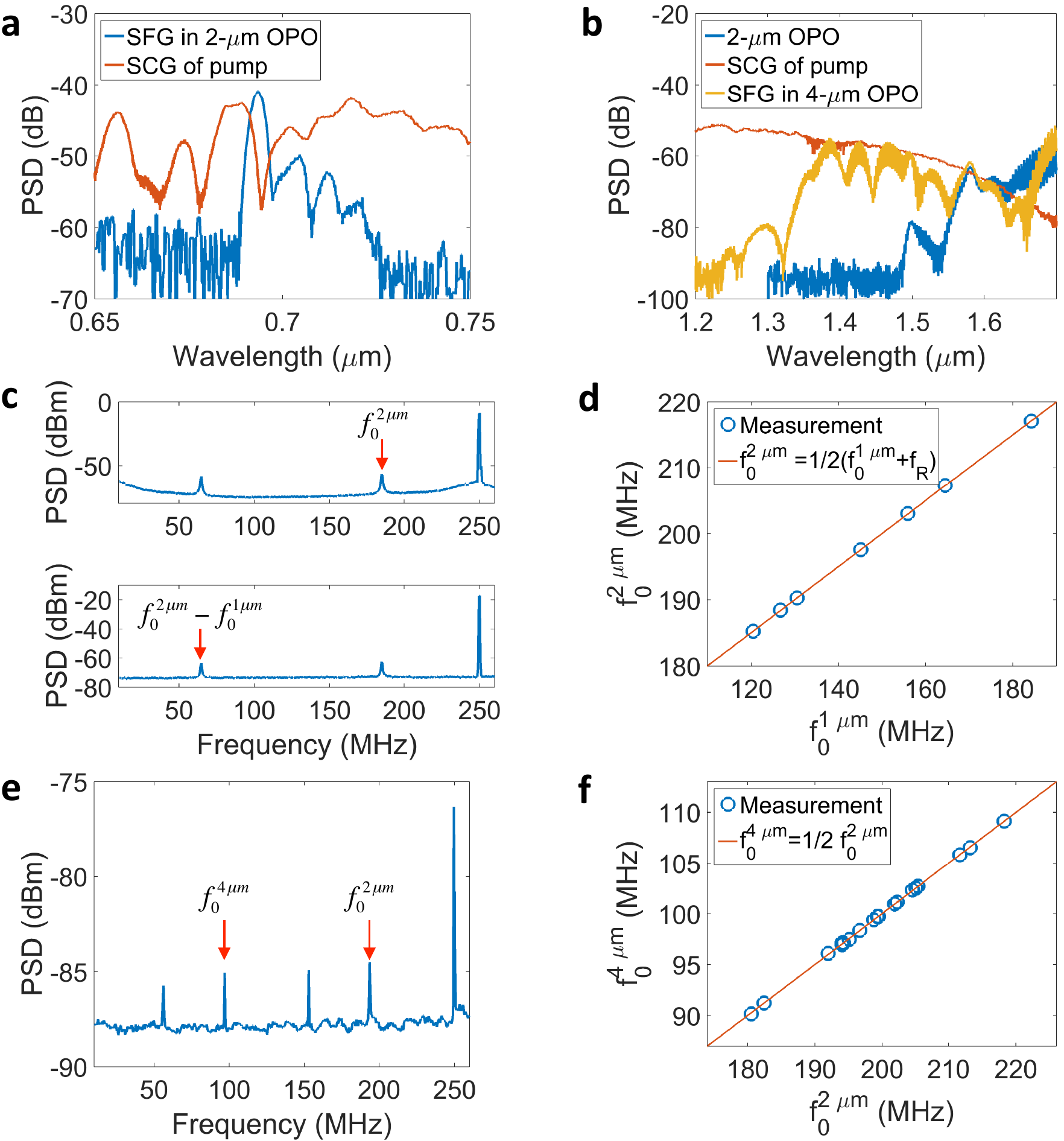}
\caption{ Beat-note measurements of the frequency combs. {\bf a}, Optical spectra around 700 nm from the SFG in the 2-$\mu$m OPO, and SCG of the pump laser. {\bf b,} Optical spectra around 1.4 $\mu$m from the blue-side of the 2-$\mu$m comb, SCG of the pump laser, and the SFG in the 4-$\mu$m OPO. {\bf c,} Sample RF spectra for measurement of $f_0^{2\mu m}$ and $f_0^{2\mu m}-f_0^{1\mu m}$. {\bf d}, CEO frequencies of the 2-$\mu$m comb as a function of pump CEO frequency. {\bf e}, Sample RF spectrum for measurement of $f_0^{2\mu m}$ and $f_0^{4\mu m}$. {\bf f}, CEO frequencies of the 4-$\mu$m comb as a function of the CEO frequency of the 2-$\mu$m comb.}
\label{fig:beatnote}
\end{figure}

To measure the CEO frequencies, we perform three heterodyne beat-note measurements. The first measurement is done at around 700 nm to measure $f_0^{2 \mu m}$. The 2-$\mu$m OPO has an output resulting from the sum frequency generation (SFG) between the 1-$\mu$m pump and the 2-$\mu$m signal which is centered around 700 nm. The CEO frequency of this output is given by:
\begin{equation}
f_{0}^{700nm}=f_{0}^{1\mu m}+f_{0}^{2\mu m}.
\end{equation}
We extend the pump spectrum through supercontinuum generation (SCG) in a tapered photonic crystal fiber (PCF) and interfere the resulting 700-nm output with the SFG signal from the 2-$\mu$m OPO. This interference is filtered using a short-pass filter at 700 nm (F$_1$) and the beat frequency, $f_0^{2\mu m}$, is measured by a silicon detector. The optical spectra of these interfering signals are shown in Fig. \ref{fig:beatnote}a.

The second beat-note, to obtain  $f_0^{2\mu m}-f_0^{1\mu m}$, is measured around 1.6 $\mu$m by interfering the red-side of the pump SCG and the blue-side of the 2-$\mu$m OPO signal output. The optical spectra of these interfering signals are shown in Fig.  \ref{fig:beatnote}b. The $f_0^{2\mu m}$ and $f_0^{2\mu m}-f_0^{1\mu m}$ are measured sequentially because of the slight difference in the required optical delays. Sample RF spectra of these interferences are shown in Fig. \ref{fig:beatnote}c, from which $f_0^{1\mu m}$ and $f_0^{2\mu m}$ are inferred by arbitrarily picking the RF peaks, and one of the two relations of eq. \ref{eq:2um} that satisfies the choice. $f_0^{1\mu m}$  is then scanned by changing the cavity configuration of the pump laser and letting it equilibrate for several minutes. The measurement is repeated for several $f_0^{1\mu m}$, and the results are shown in Fig. \ref{fig:beatnote}d, verifying the locking of the  $f_0^{2\mu m}$ to $f_0^{1\mu m}$.

To confirm that the 4-$\mu$m comb is also intrinsically locked to the 2-$\mu$m comb, we use an output of the 4-$\mu$m OPO at around 1.4 $\mu$m resulting from the SFG between the 2-$\mu$m and the 4-$\mu$m combs as shown in Fig. \ref{fig:beatnote}b. The CEO frequency of this output is:
\begin{equation}
f_{0}^{1.4 \mu m}=f_{0}^{2\mu m}+f_{0}^{4\mu m}.
\end{equation}
Since the spectrum of this output overlaps with the blue-side of the 2-$\mu$m comb, $f_{0}^{4\mu m}$ can be measured directly by an InGaAs detector and an appropriate bandpass filter (F$_4$ in Fig. \ref{fig:sch}). The output of this detector is combined with the output of the Si detector for simultaneous measurement of $f_{0}^{2\mu m}$ and $f_{0}^{4\mu m}$. A sample of the recorded RF spectrum is shown in Fig. \ref{fig:beatnote}e, and the results of the repeated measurements by scanning the pump laser CEO frequency is shown in Fig. \ref{fig:beatnote}f. These beat-note measurements confirm that the resulting femtosecond frequency combs of the cascaded half-harmonic generation at 2 $\mu$m and 4 $\mu$m are intrinsically locked to the 1-$\mu$m pump laser.

In conclusion, we demonstrate a two-stage cascaded half-harmonic generation of femtosecond frequency combs at 2 and 4 $\mu$m useful for numerous mid-IR applications. These combs are intrinsically phase- and frequency-locked to the mode-locked laser pump. The total conversion efficiency from 1 $\mu$m to 4 $\mu$m is about 11\%, which can be improved by optimizing the dispersion, loss, and output coupling of the 2$^{nd}$ OPO. Given the power scalability of 1-$\mu$m frequency combs \cite{ybcomb} and high-power operation of nonlinear materials \cite{taira}, a higher power pump and more stages of half-harmonic generation can be used to achieve phase-locked femtosecond pulses at longer wavelengths. Through half-harmonic generation, it is possible to generate significantly shorter pulses than the pump pulses; a $\sim$four-fold pulse shortening is experimentally demonstrated \cite{charlie}. The physical size of the half-harmonic OPOs can be made very compact either using fiber-feedback cavities \cite{fiberfeedback} or harmonic-pumping \cite{fractional}. These properties make this technique a simple and universal tool for frequency-locked down conversion of frequency combs. 

In principle, the conversion efficiency of half-harmonic generation can get close to 100\%, however, for further optimization of these systems to significantly surpass the demonstrated 64\% for the 2-$\mu$m OPO, numerical simulations are required to take into account the complex nonlinear interactions of these extremely short pulses. With recent advances in the development of on-chip resonators in materials with strong second-order nonlinear susceptibility \cite{gap,gaas}, it is expected that half-harmonic generation and cascading it can potentially be implemented on-chip for the growing demand for mid-IR sources.

The authors thank M.M. Fejer,  for discussions, and R. McCracken, C. Langrock, and K. Urbanek for experimental support.

\end{document}